\documentclass{emulateapj}

\begin{document}
\def\kms{\hbox{km s$^{-1}$}}

\title{ Velocity Dispersion Profile of the Milky Way Halo }

\author{Warren R.\ Brown,
	Margaret J.\ Geller,
	Scott J.\ Kenyon}

\affil{Smithsonian Astrophysical Observatory, 60 Garden St, Cambridge, MA 02138, USA} 
\email{wbrown@cfa.harvard.edu}

\author{\and Antonaldo Diaferio}

\affil{Dipartimento di Fisica Generale `Amedeo Avogadro', Universit\`{a} degli Studi di Torino, Via P. Giuria 1, I-10125 Torino, Italy}
\affil{Istituto Nazionale di Fisica Nucleare (INFN), Sezione di Torino, Via P.\ Giuria 1, I-10125, Torino, Italy}

\shorttitle{ Velocity Dispersion Profile }
\shortauthors{Brown et al.}

\begin{abstract}
	We present a spectroscopic sample of 910 distant halo stars from the
Hypervelocity Star survey from which we derive the velocity dispersion profile of
the Milky Way halo.  The sample is a mix of 74\% evolved horizontal branch stars and
26\% blue stragglers.  We estimate distances to the stars using observed colors,
metallicities, and stellar evolution tracks.  Our sample contains twice as many
objects with $R>50$ kpc as previous surveys.  We compute the velocity dispersion
profile in two ways:  with a parametric method based on a Milky Way potential model,
and with a non-parametric method based on the caustic technique originally developed
to measure galaxy cluster mass profiles.  The resulting velocity dispersion
profiles are remarkably consistent with those found by two independent surveys based
on other stellar populations:  the Milky Way halo exhibits a mean decline in radial
velocity dispersion of $-0.38\pm0.12$ \kms\ kpc$^{-1}$ over $15<R<75$ kpc.  This
measurement is a useful basis for calculating the total mass and mass distribution
of the Milky Way halo.
	\end{abstract}


\clearpage
\section{INTRODUCTION}

	A fundamental property of the Milky Way galaxy is its total mass.  In the
Cold Dark Matter paradigm, the total mass of a galaxy's dark matter halo correlates
with its merger history, star formation history, and number of its satellite
sub-halos.  A galaxy's mass distribution is also a fundamental constraint on
theories.  Cold dark matter models predict that density follows an universal NFW
profile \citep{nfw97}; in Modified Newtonian Dynamics there is no dark matter halo
and the mass distribution is highly flattened compared to cold dark matter models
\citep{hernandez09}.  The Milky Way provides an unique laboratory for testing these
issues.

	Nevertheless, the mass and mass distribution of the Milky Way are among the
most poorly known of all Galactic parameters, especially at large distances $R
\gtrsim 50$ kpc.  Total mass estimates span at least a factor of 4, from
$0.5\times10^{12}$ to $2\times10^{12}$ M$_{\odot}$ (see below).  Recent theoretical
work based on semi-analytic models and timing arguments suggest the Milky Way may
be even more massive \citep{li08, abadi09}.  Clearly, the mass and mass distribution
of the Milky Way remain controversial and important issues.

	A powerful approach for measuring the Milky Way's mass distribution is
measuring the motions of tracer objects.  Radio masers probe the Galactic rotation
curve out to 15 kpc \citep{reid09}; H {\sc i} gas probes the rotation curve to
larger distances \citep{kalberla08}.  Luminous tracers at $R \gtrsim 50$ kpc,
however, are rare. Historically, globular clusters and dwarf galaxies were used to
measure the total mass of the Milky Way \citep[e.g.][]{little87, zaritsky89,
kulessa92, kochanek96, wilkinson99}.  These mass estimates were based on samples of
a few dozen objects, and suffered from a systematic factor of $\sim$2 uncertainty
depending on the inclusion of Leo I.  Post-main sequence halo stars provide denser
tracers, but the blue horizontal branch (BHB) star samples of \citet{sommer97} and
\citet{sakamoto03} are limited to $R\lesssim20$ kpc.  \citet{battaglia05} added 58
distant red giants from the Spaghetti survey and claimed a large decline in the
velocity dispersion at $R>$50 kpc.  Recently, \citet{xue08} analyzed the Sloan
Digital Sky Survey (SDSS) sample of 2,466 BHB stars and found a small decline in
velocity dispersion with distance.  Only 80 of the SDSS BHB stars are located at
$R>50$ kpc.

	Here we use the distant halo stars from the hypervelocity star (HVS) program
\citep{brown05, brown06, brown06b, brown07a, brown07b, brown09a, brown09b} to
measure the velocity dispersion profile of the Milky Way.  Our dataset is a complete
spectroscopic sample of 910 stars observed over 7300 deg$^2$ of the SDSS Data
Release 6 imaging region.  We gain increased leverage on the velocity dispersion
profile for $R \gtrsim 50$ kpc.  We find an average decline in radial velocity
dispersion of $0.38\pm0.12$ \kms\ kpc$^{-1}$ over $15<R<75$ kpc.

	In \S 2 we describe the observations and luminosity estimates for stars in
our sample.  In \S 3 we make parametric and non-parametric estimates of the velocity
dispersion profile.  We also discuss possible systematics, and compare our results
with earlier work.  We conclude in \S 4.  The data are in the Appendix.

\section{DATA}

	The hypervelocity star program is a radial velocity survey of stars selected
with the colors of late B-type stars.  The radial velocity survey is now 93\%
complete over 7300 deg$^2$ of the SDSS DR6 imaging footprint.  Here we focus
exclusively on the stars with late-B and early-A spectral types; we exclude all
white dwarfs \citep{kilic07, kilic07b}, B supergiants \citep{brown07c}, emission
line galaxies \citep{kewley07, brown08a}, and quasars \citep{brown09a}.

	The dataset contains 910 stars:  571 stars from the original HVS survey
\citep{brown07b}, 331 stars from the new HVS survey \citep{brown09a}, and 8 BHB
stars from the earliest sample \citep{brown05}.  We begin by describing the
observables -- magnitude, position, velocity -- all of which are well-determined.  
Stellar luminosity is less well-determined and must be inferred from colors and
metallicity.  We discuss the luminosity estimates and distance determinations in
some detail.
	The observed and derived quantities for each star are listed in Table
\ref{tab:dat}, described in the Appendix.

\subsection{Photometry}

	All photometry comes from SDSS Data Release 6 \citep[][]{adelman08}.  We use
uber-calibrated PSF magnitudes, and correct the magnitudes and colors for reddening
following \citet{schlegel98}.

\begin{figure}		
 \plotone{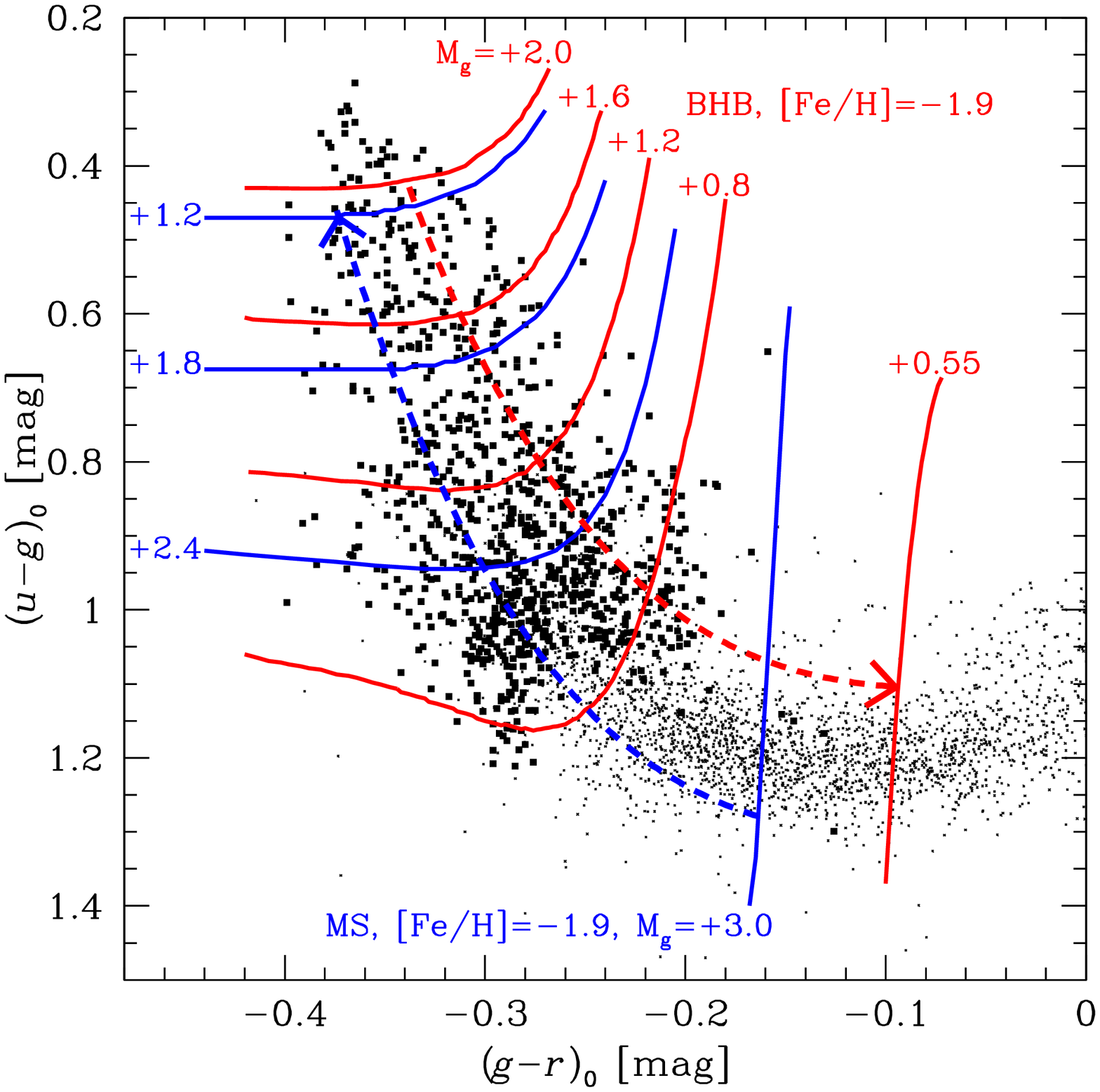}
 \caption{ \label{fig:ugr}
	Color-color diagram showing the distribution of the HVS survey stars ({\it
solid squares}) and the \citet{xue08} BHB stars ({\it dots}) compared to lines
of constant absolute magnitude $M_g$ for BHB stars \citep[red lines,][]{dotter08}
and main sequence stars \citep[blue lines][]{girardi04}.  All tracks are for
[Fe/H]=$-1.9$, the mean metallicity of halo stars.  Dashed lines indicate the
direction of increasing luminosity.  At [Fe/H]=$-1.9$, BHB stars and blue stragglers
share identical luminosities around $(u-g)_0\simeq0.6$. }
 \end{figure}

\subsection{Target Selection}

	The HVS survey target selection emphasizes outliers in the halo population:  
we target stars redder in $(u-g)_0$ than known white dwarfs and bluer in $(g-r)_0$
than known BHB stars \citep[][]{brown06b}.  This color cut through the stellar
population, illustrated in Figure \ref{fig:ugr}, allows us to detect hypervelocity
stars efficiently.  The majority of targets, however, are normal halo stars.

	Our target selection includes stars with $17<g_0<19.5$ in the range
$-0.39<(g-r)_0<-0.25$ \citep[][]{brown07b} and fainter stars with $19<g_0<20.5$ over
a broader color range $-0.40<(g-r)_0<-0.20$ \citep{brown09a}.  We also include 8
confirmed BHB stars from \citet{brown05} with $19.5<g_0<20.25$ and
$-0.3<(g-r)_0<-0.1$.  All 910 targets are located in the SDSS DR6 footprint and have
an average surface density on the sky of 0.12 deg$^{-2}$.

\begin{figure}		
 \plotone{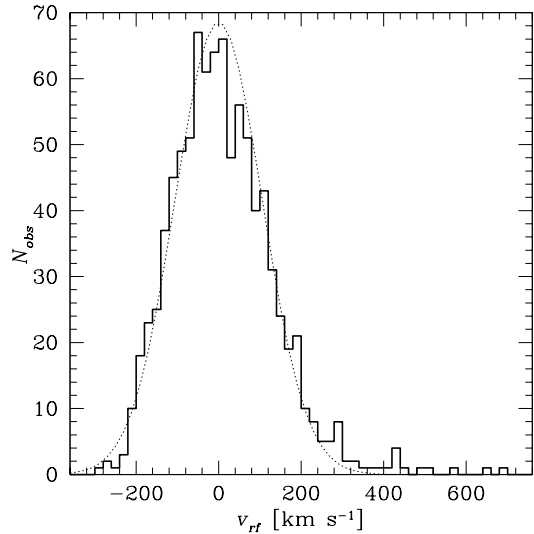}
 \caption{ \label{fig:velh}
	Distribution of velocities in the Galactocentric rest frame.  The dotted 
line shows a fiducial Gaussian with zero mean and 106 \kms\ dispersion.  The 
positive velocity outliers with $v_{rf}>+400$ \kms\ are the unbound HVSs.}
 \end{figure}

\subsection{Radial Velocity}

	We obtained spectroscopic observations at the 6.5m MMT telescope with the
Blue Channel spectrograph.  We operated the spectrograph with the 832 line mm$^{-1}$
grating in second order, providing wavelength coverage 3650 \AA\ to 4500 \AA\ and a
spectral resolution of 1.2 \AA.  We obtained all observations at the parallactic
angle, with a comparison lamp exposure for every survey object.

	We processed the data using IRAF\footnote{IRAF is distributed by the
National Optical Astronomy Observatories, which are operated by the Association of
Universities for Research in Astronomy, Inc., under cooperative agreement with the
National Science Foundation.} in the standard way.  We measure radial velocities by
cross-correlating the observations with radial velocity standards \citep{fekel99}
using the package RVSAO \citep{kurtz98}.  The average radial velocity uncertainty of
the stars is $\pm12$ km s$^{-1}$.

	All velocities discussed here are in the Galactocentric rest frame,
indicated $v_{rf}$.  Given the outer halo location of the stars, we note that the
observed radial velocities are almost purely ($>$85\%) radial in the Galactocentric
frame.  We transform heliocentric velocities ($v_{helio}$) into Galactocentric rest
frame velocities assuming a circular velocity of 220 \kms\ and a solar motion of
($U, V, W$) = (10, 5.2, 7.2) \kms\ \citep{dehnen98}:
	\begin{equation} \label{eqn:vrf}
v_{rf} = v_{helio} + 220\sin{l}\cos{b} + (10\cos{l}\cos{b} + 5.2\sin{l}\cos{b} + 7.2\sin{b}).
	\end{equation} 
	\citet{reid09} argue for a larger circular velocity of 250 \kms\ based on
trigonometric parallaxes to star formation regions in the disk.  We test using a
circular velocity of 250 \kms\ and find statistically identical velocity dispersion
profiles.  This insensitivity to circular velocity arises because the stars in our
high latitude survey have a mean value of $|\sin{l}\cos{b}|=0.32$; changing the
Sun's circular velocity by 30 \kms\ results in a $\pm$10 \kms\ change in the stars'
rest frame velocities.  10 \kms\ is smaller than our measurement error and an order
of magnitude smaller than the velocity dispersion of the stars.  \citet{mcmillan09}
show that the \citet{reid09} data are consistent, at the 1-$\sigma$ level, with the
canonical circular velocity of 220 \kms .  Thus we use 220 \kms\ here.

	Figure \ref{fig:velh} plots the resulting Galactic rest frame velocity
distribution of the 910 halo stars.  For reference, we also draw a Gaussian with
zero mean and 106 \kms\ dispersion (dotted line).  The observations reveal a
significant asymmetry of outliers in the tails of the velocity distribution.  There
are no stars with $v_{rf}<-300$ \kms\ but 18 stars with $v_{rf}>+300$ \kms (the
HVSs).  We address this issue in Section \ref{sec:vd}.

\subsection{Metallicity}

\begin{figure}		
 \plotone{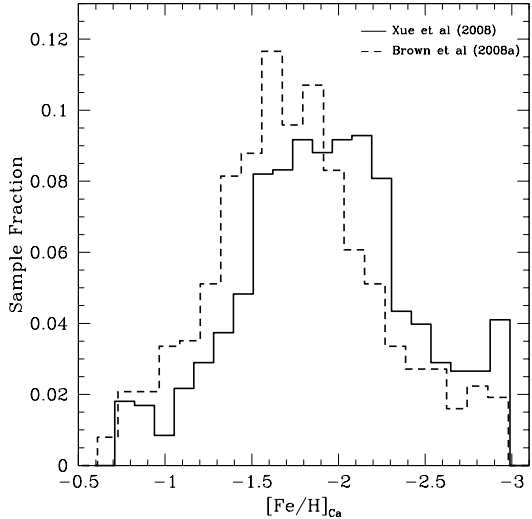}
 \caption{ \label{fig:feh}
	Metallicity distribution function of inner halo BHB stars \citep{brown08b}
and outer halo BHB stars \citep{xue08} based on Ca {\sc ii} K.  The shift towards
lower metallicity in the outer halo is expected; the \citet{xue08} BHB stars with
$g_0>17$ have a mean [Fe/H]$_{Ca}=-1.9$. }
 \end{figure}

	The strongest metal line in our spectra is the 3933 \AA\ Ca {\sc ii} K line.  
Unfortunately, at the effective temperatures sampled by the survey, $10,000 \le
T_{eff} \le 15,000$ K, the equivalent width of Ca {\sc ii} K is small
\citep{wilhelm99a}.  Thus, Ca {\sc ii} K provides poor leverage on the metallicity
of our stars.  Metallicity is better determined for redder (cooler) BHB stars, for
example the BHB survey of \citet{brown08b} and the BHB sample from the SDSS
spectroscopic survey \citep{xue08}.  The metallicity distributions of these two BHB
samples are plotted in Figure \ref{fig:feh} and are similar in shape.  However, the
outer halo sample of \citet{xue08} is about 0.2 dex more metal-poor than the inner
halo sample of \citet{brown08b}.  The mean metallicity of the \citet{xue08} BHB
stars with $g_0>17$ is [Fe/H]$_{Ca}=-1.9$.

	To make luminosity estimates, we assume our survey stars have the
metallicity distribution function of \citet{xue08}.  This assumption is reasonable
given the very similar sky coverage and penetration into the halo of the two
surveys: the stars occupy similar regions of the Milky Way halo.  The reddest stars
in our sample, where we can estimate metallicity, are metal-poor \citep{brown06b},
consistent with the metallicity distribution function of \citet{xue08}.

\subsection{Spectroscopic Identification}

	Although the stars in our survey have the spectral types of late B- and
early A-type stars, their nature is ambiguous.  The old stellar population of the
halo contains both evolved BHB stars and main sequence blue stragglers.  Halo
surveys consistently find that $\sim$50\% of A-type stars in the field are blue
stragglers \citep{norris91, kinman94, preston94, wilhelm99b, clewley02, clewley04,
brown03, brown05b, brown08b, xue08}.  This result is problematic because BHB stars
and blue stragglers can have very different luminosities (Figure \ref{fig:ugr}).  
Spectroscopic measures of surface gravity can discriminate the evolutionary state of
the stars \citep{kinman94, wilhelm99a, clewley02, clewley04}.  Unfortunately,
surface gravity measures fail for our sample because BHB and main sequence stars
have nearly identical surface gravities at the effective temperatures of our stars.

	We target stars so blue, however, that we largely exclude the possibility of
blue stragglers.  A $\simeq$0.75 M$_{\odot}$ star with [Fe/H]=-1.9 has a main
sequence lifetime of a Hubble time \citep{girardi04}.  A field blue straggler cannot
plausibly have more than twice the mass of a main-sequence turnoff star.  In our
sample, 36\% of the stars are bluer than both $(g-r)_0=-0.27$ and $(u-g)_0=0.87$,
the color of a 1.5 M$_{\odot}$ star with [Fe/H]=-1.9 \citep{girardi04}.  Thus the
nature of many of the stars is clear:  they are hot BHB stars.  

	Stars bluer than a 1.5 M$_{\odot}$ star are BHB; stars redder than a 1.5
M$_{\odot}$ star are equally likely to be BHB stars or blue stragglers.  We base
this conclusion on \citet{xue08}, who observe a BHB fraction of 47\% at the red end
of our sample.  Overall, our sample is 74\% BHB stars and 26\% blue stragglers.

\subsection{Luminosity}

	We estimate luminosity from stellar evolution tracks.  \citet{girardi02,
girardi04} provide main sequence tracks in the SDSS passbands for metallicities
ranging from solar to [Fe/H]=$-2.3$, but do not include the horizontal branch.  
\citet{dotter07, dotter08} provide BHB tracks in the SDSS passbands over the same
range of metallicities.  To simplify the luminosity estimates, we fit low-order
polynomials to the tracks as a function of both color and metallicity.

\begin{figure}		
 \plotone{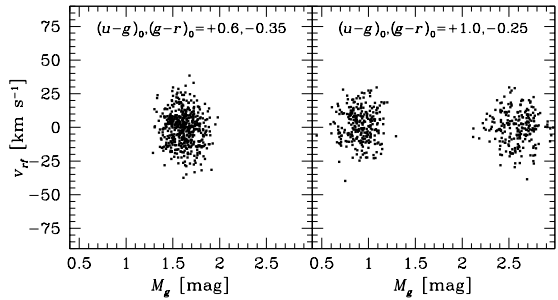}
 \caption{ \label{fig:mvplot}
	Distribution of luminosity $M_g$ and velocity for fiducial stars in the
bluest (left panel) and reddest (right panel) quartiles of our sample.  The
distributions are obtained by propagating the observed uncertainties of a $g_0=19$
star at the median depth of our survey ($\sigma_{u-g}=0.05$, $\sigma_{g-r}=0.03$,
$\sigma_v=12$ km s$^{-1}$) with the \citet{xue08} halo metallicity distribution
function through stellar evolutionary tracks for main sequence \citep{girardi04} and
BHB \citep{dotter08} stars.  Blue stars have luminosity precise to 10\%, whereas red
stars have a bimodal distribution in luminosity (compare with Figure
\ref{fig:ugr}).}
 \end{figure}

	We estimate luminosity two ways, using a star's $(u-g)_0$ and $(g-r)_0$
color.  Our final luminosity is the weighted average of the two estimates.  We
weight by the slope of the observed $(u-g)_0$ versus $(g-r)_0$ distribution (Figure
\ref{fig:ugr}). This weighting is important because luminosity is most sensitive to
$(u-g)_0$ at the blue end of our sample and to $(g-r)_0$ at the red end of our
sample.  The weighting favors luminosities derived from $(u-g)_0$ for
$(g-r)_0<-0.24$ and luminosities derived from $(g-r)_0$ for $(g-r)_0>-0.24$.  
Figure \ref{fig:ugr} shows the resulting lines of constant luminosity for a star
with [Fe/H]=$-1.9$ from the \citet{girardi04} main sequence tracks (in blue) and
from the \citet{dotter08} BHB tracks (in red).

	The accuracy of the luminosity estimate depends on the accuracy of the
stellar evolution models.  For example, alpha-enhanced tracks (appropriate for the
old stellar halo populations) are systematically bluer than solar-scaled tracks for
both main sequence and BHB stars \citep{lee09}.

	The precision of the luminosity estimate is more robust because we apply the
same tracks to all stars.  The precision depends on observational uncertainties and
is straightforward to quantify.  A $g_0=19$ star at the median depth of our survey
has uncertainties $\sigma_{(u-g)}=0.05$ and $\sigma_{(g-r)}=0.03$ in color and
$\sigma_v=12$ km s$^{-1}$ in velocity.  We assume that metallicity is randomly drawn
from the \citet{xue08} metallicity distribution function (Figure \ref{fig:feh}). We
then propagate these uncertainties through the main sequence and BHB tracks to
visualize the resulting distribution of luminosity estimates.

	Figure \ref{fig:mvplot} plots the distribution of luminosity and velocity
(centered at zero) for two fiducial stars in the bluest quartile (left panel) and
the reddest quartile (right panel) of the survey.  Stars in the bluest quartile have
luminosities precise to 10\% because main sequence and BHB stars have essentially
identical luminosities at these colors (Figure \ref{fig:ugr}).  Stars in the reddest
quartile, however, have bimodal luminosity distributions with a factor of 4 spread
in luminosity (a factor of 2 in distance).  This degeneracy is broken for confirmed
BHB stars:  33 of our stars are listed as BHB stars in \citet{xue08} and 8 are BHB
stars from our earliest sample \citep{brown05}.  For these confirmed BHB stars we
use only the BHB luminosity to calculate distance.

	Three of the stars with $v_{rf}>+400$ \kms\ in this survey are confirmed
young main sequence B stars \citep{fuentes06, bonanos08, lopezmorales08,
przybilla08, przybilla08b}.  For the purposes of this paper, however, we treat these
HVSs the same as the other stars in our sample:  we estimate distances to the HVSs
as if they were halo BHB stars or metal-poor blue stragglers.  We clip the HVSs from
the sample when calculating the velocity dispersion in Section \ref{sec:vd}.

\begin{figure}		
 \plotone{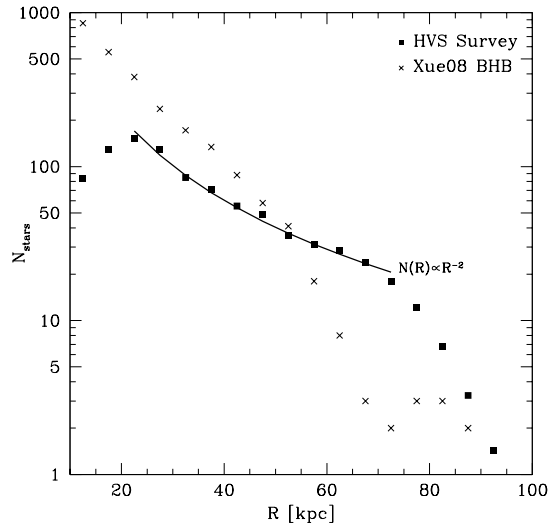}
 \caption{ \label{fig:rplot}
	Density distribution of the HVS survey stars compared with the \citet{xue08}
BHB sample.  Our survey is complete, but we find an artificially shallow density
profile because we observe over a broader color range at large depths $g_0>19$ mag.}
 \end{figure}

\subsection{Distance}

	We calculate distances from the observed apparent magnitudes and 
estimated luminosities:
 \begin{equation} d = 10^{(g_0 - M_g)/5 -2} ~{\rm kpc},\end{equation}
	where $d$ is the heliocentric distance in kpc, $g_0$ is the apparent
magnitude corrected for extinction, and $M_g$ is the absolute magnitude estimated
above.  We convert heliocentric distance $d$ to Galactocentric distance $R$ with
the Sun at $R=8$ kpc.

	Figure \ref{fig:rplot} plots the resulting distance distribution of the
survey stars compared with the \citet{xue08} sample of BHB stars.  The \citet{xue08}
sample contains more than twice as many stars as our sample; however, our sample is
deeper and contains twice as many stars with $R\gtrsim50$ kpc.  Neither sample
fairly measures the density profile of the halo.  The \citet{xue08} sample is
incomplete in color, magnitude depth, and spatial coverage.  Our sample is complete
in all dimensions, but we find an artificially shallow density profile -- consistent
with $N(R)\propto R^{-2}$ over the range $20<R<70$ kpc (Figure \ref{fig:rplot}) --
because we observe over a broader color range starting at $g_0>19$ mag.  A shallow
density profile works to our advantage for the velocity dispersion analysis,
however, because our stars sample greater distances at a greater relative density.

\section{HALO VELOCITY DISPERSION PROFILE \label{sec:vd}}

	Based on the distances and velocities of the sample stars, we calculate the
velocity dispersion profile of the Milky Way halo.  We use two independent methods
to calculate velocity dispersion: a parametric method based on a Milky Way potential
model, and a non-parametric method based on the \citet{diaferio97} caustic
technique, originally developed to measure galaxy cluster mass profiles.  Using
these two methods provides a measure of the systematic error introduced from
clipping velocity outliers.  We also discuss the systematic error from binary stars
and disk stars in our dataset, and we conclude by comparing our velocity dispersion
profiles with previous work.

\subsection{Computational Approach}

	We use a Monte Carlo approach to model the distance to each star.  We assume
that photometric and velocity errors are Gaussian distributed, and that the
underlying metallicity distribution is that of \citet{xue08}.  We then derive the
luminosity for each star by randomly drawing its color and metallicity from the
observed distributions and comparing them to the main sequence and BHB tracks (Fig.\
\ref{fig:ugr}) described above.  We sample the distributions 100 times per star.  
Using a Monte Carlo approach allows us to account for the non-Gaussian distribution
of luminosity estimates unique to each star.

\begin{figure}		
 \plotone{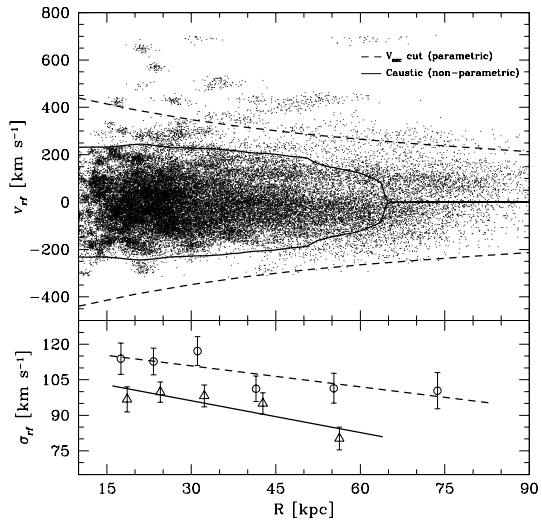}
 \caption{ \label{fig:vd}
	Upper panel:  Monte Carlo distribution of distance and radial velocity for
our sample. We represent each star by 100 random draws from its parent distance and
velocity error distribution.  We clip velocity outliers using two approaches:  a
parametric method based on a Milky Way potential model ({\it dashed line}), and a
non-parametric method based on the \citet{diaferio97} caustic technique ({\it solid
line}).
	Lower panel:  binned velocity dispersion profile from the two approaches.  
Errors are from bootstrap re-sampling.  Linear least squares fits show an average
$-0.38\pm0.12$ \kms\ kpc$^{-1}$ decline in velocity dispersion over $15<R<75$ kpc.}
 \end{figure}

	The resulting Monte Carlo catalog drawn from the observations produces the 
``cloud'' of velocities and distances shown in Figure \ref{fig:vd}; the
distribution of points reflects the uncertainties in the velocity measurements and
the distance estimates of the stars.

	We calculate the velocity dispersion profile by grouping stars into 5 or 6
unique bins in distance.  The bins have sizes of 0.33$R$, chosen so that the bins
contain at least 70 stars.  We require 50 stars to obtain a dispersion with a formal
statistical uncertainty of 10\%.  Bins with smaller occupation have dispersions
systematically biased low resulting from small number statistics.  Given the
distance distribution of our sample, the occupation requirement constrains our
velocity dispersion measurements to the region $15<R<75$ kpc.
	Stars with well-determined luminosities contribute their full weight to a
single distance bin; stars with poorly-constrained luminosities contribute less
weight distributed over multiple bins.

	We use bootstrap re-sampling to calculate the uncertainty in the velocity
dispersion measurement.  Our procedure is to draw random sets of 910 stars, with
replacement, from the Monte Carlo catalog and re-calculate the velocity dispersion.  
We re-sample 10,000 times; the uncertainty is the standard deviation of the velocity
dispersions.

\subsection{Velocity Dispersion: Potential Model}

	Our survey was designed to find unbound HVSs.  Thus the first step in
calculating the velocity dispersion profile is to clip the unbound stars from the
sample.  Defining an unbound star is difficult, however.  The best estimate of the
Galactic escape velocity is $550\pm50$ \kms\ at the solar circle, based on the
highest velocity stars in the solar neighborhood \citep{smith07}.  The escape
velocity of the outer halo is more poorly constrained; extrapolating $v_{esc}(R)$ to
large $R$ requires a potential model.

	Here, we use the \citet{kenyon08} potential model that is tied to observed
Milky Way mass measurements.  To establish $v_{esc}(R)$ in this model, we drop a
test particle at rest from $R=500$ kpc (i.e., halfway to M31) and calculate its
trajectory down to $R=0$ kpc.  The \citet{kenyon08} model predicts $v_{esc}=585$
\kms\ at $R=8$ kpc in this definition of escape velocity, in good agreement with the
\citet{smith07} observation.  We fit a low-order polynomial to the calculated
velocity as a function of distance and find
	\begin{equation} \label{eqn:esc}
 v_{esc}(R) = -2.30\times10^{-4} R^3 + 0.0588 R^2 -6.62 R + Z ~~{\rm km~s^{-1}},
	\end{equation} where $Z=619$ \kms , valid for $15<R<100$ kpc.  

	To avoid imposing an arbitrary mass on our velocity dispersion measurement,
however, we re-normalize Equation \ref{eqn:esc} to the observed envelope of negative
velocity stars in our sample.  Figure \ref{fig:vd} shows the $v_{esc}(R)$ relation
that we use, with $Z=500$ \kms\ (Eqn.\ \ref{eqn:esc}).  Our assumption in changing
$Z$ to 500 \kms\ is that the velocity distribution of halo stars extends up to the
escape velocity, and that the most negative velocity stars -- the stars falling in
from the largest distances -- provide the most robust measure of escape velocity in
our dataset.  \citet{perets09b} discuss this further in the context of HVSs.  We 
test the effects of our choice of $v_{esc}(R)$ by using a non-parametric approach to
calculate velocity dispersion in the next section.  In this section, we calculate
the velocity dispersion profile by using the shape of the $v_{esc}(R)$ relation to
provide a physically motivated means of clipping velocity outliers.

	We plot the resulting velocity dispersion of stars with velocities less than
$v_{esc}(R)$ for $Z=500$ \kms\ in the lower panel of Figure \ref{fig:vd}.  
Our Monte Carlo approach allows outliers that might otherwise be clipped to
contribute a weight appropriate to their measurement uncertainties. A linear least
squares fit to the velocity dispersion profile finds a declining velocity
dispersion,
	\begin{equation}
 \sigma_v = (-0.30 \pm 0.10)R + (120\pm4.6) ~{\rm km~s^{-1}},
	\end{equation} valid over $15<R<75$ kpc.  A higher-order fit does not
significantly improve the residuals.  The linear fit has a standard deviation of 4.8
\kms\ and a reduced $\chi^2$ of 0.7.  A fixed velocity dispersion, by comparison,
has a standard deviation of 7.6 \kms\ and a reduced $\chi^2$ of 1.5, poorer than the
linear fit but not statistically inconsistent.

\subsection{Velocity Dispersion:  Caustic Method}

	A non-parametric approach to determining the velocity dispersion profile
leads to robust results based on fewer a-priori assumptions.  Here we use the
caustic technique.  Because this technique is new to the stellar halo literature, we
begin with an introduction before discussing the mechanics and results.
\citet{diaferio09} provides a recent review of the technique.  

	The caustic technique was originally developed to measure galaxy cluster
mass profiles \citep{diaferio97, diaferio99}.  Caustics are essentially escape
velocity curves.  The caustic technique is based on the distribution of objects in
phase space: the line-of-sight velocity $v$ vs.\ projected distance $r$ from the
cluster center.  In this plane, cluster members distribute in a characteristic
trumpet shape with upper and lower borders indicating the escape velocity from the
system.  The amplitude ${\cal A}(r)$ of this distribution, which decreases with $r$,
is a direct measure of the escape velocity from the system, independent of its
dynamical state.  The caustic technique only assumes that the spatial distribution
of tracers is spherically symmetric. 

	The caustic technique can be applied to any self-gravitating system.  The
technique is non-parametric and does not require that the tracers of the potential
be in virial equilibrium.  It is an effective tool for identifying bound members of
the system defined by the upper and lower caustics (sharp declines in phase space
density of tracers).  For example, \citet{serra09} apply the caustic technique to
five dwarf spheroidals of the Milky Way.

	Applying the caustic technique to the halo star sample requires a slight
modification of the procedure used for galaxy clusters.  Traditionally, the
technique arranges sample objects in a binary tree according to their pairwise
projected binding energies \citep{diaferio99}. By walking along the main branch of
the tree, the technique determines the velocity dispersion which locates the caustic
in the velocity diagram. This step does not apply to the Milky Way halo; the stars
are not obviously clustered on the sky and the estimate of the pairwise binding
energy is inappropriate.  Thus we apply the technique directly to the phase space
diagram shown in Figure \ref{fig:vd}.

	The caustics are the curves satisfying the equation $f_q(r,v)=\kappa$, where
$f_q(r,v)$ is the distribution in the phase space diagram, and $\kappa$ is the
root of the equation $\langle v_{\rm esc}^2\rangle_{\kappa,R}=4\sigma^2$. The
function \begin{equation}
	\langle v_{\rm esc}^2\rangle_{\kappa,R}=\int_0^R{\cal
A}_\kappa^2(r)\varphi(r)dr/ \int_0^R\varphi(r)dr
	\end{equation} is the mean caustic amplitude within $R$, $\varphi(r)=\int
f_q(r,v) dv$.  $R$ is not a free parameter in the standard application of the
technique, but here we chose $R=30$~kpc. Our results are totally insensitive to this
parameter: varying $R$ in the range 20-100~kpc varies the final number of the halo
members by at most five and does not change the velocity dispersion profile.

	We use an iterative approach to estimate the velocity dispersion profile.
First we compute the velocity dispersion of the total sample and locate the caustics
that enable some interloper removal. We then compute a new velocity dispersion with
the member stars, locate new caustics and remove further interlopers. We proceed
until no star is identified as an interloper. This procedure requires only four
steps and on average yields 838 (of 910) final halo members and an overall velocity
dispersion of 99~km~s$^{-1}$.  The final set of caustics are drawn with the solid
line in the upper panel of Figure \ref{fig:vd}.

	The open triangles in the lower panel of Figure \ref{fig:vd} show the
caustic velocity dispersion profile.  It is visually apparent that the caustic
technique measures the velocity dispersion from the well-sampled core of the
velocity distribution.  This approach results in a reliable velocity dispersion
profile, but yields a velocity dispersion that is systematically 10\% smaller than
found by our parametric $v_{esc}(R)$ method.  We conclude that clipping velocity
outliers with the caustic technique changes the observed amplitude, but not the
observed slope, of the velocity dispersion profile.

	A linear least squares fit to the caustic velocity dispersion profile finds 
a declining velocity dispersion,
	\begin{equation}
	\sigma_v=(-0.45 \pm 0.16) R + (110\pm6.0) ~{\rm km~s^{-1}},
	\end{equation} valid over $16<R<64$ kpc.  The linear fit has a standard
deviation of 4.8 \kms\ and a reduced $\chi^2$ of 1.0.  A fixed velocity dispersion,
by comparison, has a standard deviation of 7.9 \kms\ and a reduced $\chi^2$ of 2.8.  
In this case, a constant velocity dispersion is a significantly poorer fit to the
data.

\subsection{Possible Systematics}

	There are at least two contaminants that may systematically affect our
velocity dispersion profile:  binary stars, which increase the velocity
dispersion, and disk stars (i.e.\ white dwarfs), which decrease the velocity
dispersion.  We investigate how these contaminants may alter the observed velocity
dispersion profile.

	Binary stars are unlikely to change the velocity dispersion profile.  Both
BHB stars and 1.5 M$_{\odot}$ blue stragglers have stellar radii around 2.5
R$_{\odot}$.  Equal mass pairs of such stars must have semimajor axes of at least
6.5 R$_{\odot}$ to avoid Roche lobe overflow.  Thus the most compact possible binary
system has velocity semi-amplitude of 100 \kms .  Assuming that half of the targets
are binaries with a log-normal distribution of semimajor axes, we expect 30 binaries
in our dataset with $v\sin{i}>$50 \kms .  This estimate is generous given that BHB
stars have recently evolved through the red giant phase and are thus unlikely to
have close companions.  In any case, binaries will be observed at a random orbital
phase and binned with 70-200 other stars to compute a velocity dispersion.  We
propagate our simulated distribution of $v\sin{i}$'s through our Monte Carlo catalog
and find a negligible change in the velocity dispersion profile; the uncertainty in
the velocity dispersion slope remains 30\%.

	Disk stars pose a greater threat to the velocity dispersion profile.  Disk
stars have a systematically lower velocity dispersion than halo stars and may also
have a non-zero mean velocity because of the longitude dependence of the Sun's
circular velocity correction.  The SDSS imaging survey from which we draw our
candidates does not uniformly survey the sky across all longitudes.


	Our greatest concern is white dwarfs, which are intrinsically faint and
overwhelmingly appear at faint magnitudes in our survey.  Inserting white dwarfs
into our halo sample thus reduces the velocity dispersion observed at large
distances.  Observationally, 15\% of our survey targets are white dwarfs, mostly
found at low latitudes.  If we insert a 1\% white dwarf contamination into our Monte
Carlo catalog, the resulting velocity dispersion profile steepens by 10\%.  
Fortunately, white dwarfs are readily identified by their broad Balmer lines
\citep[see][]{kilic07}.  Visual inspection of our spectra reveals no white dwarf
contaminants.

	Binaries also have observational constraints.  We obtained repeat
observations for every star with $|v_{rf}|>300$ \kms\ and found only one star, a
star near $-$300 \kms , to exhibit radial velocity variation.  We thus excluded this
star from the halo sample.  We conclude that binaries and white dwarfs are unlikely
to significantly influence the velocity dispersion profile measured from our halo
star sample.

\begin{figure}		
 \plotone{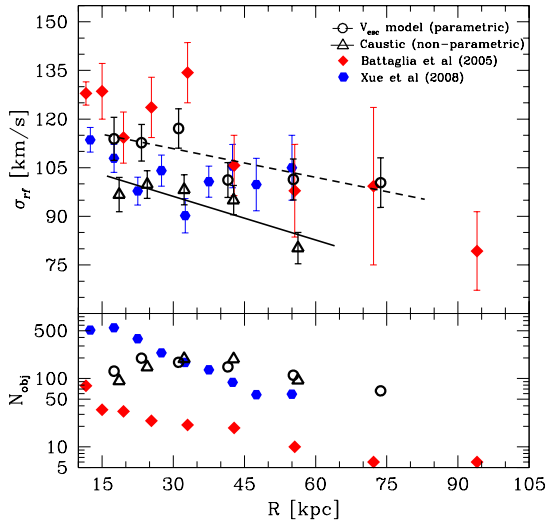}
 \caption{ \label{fig:compare}
	Milky Way velocity dispersion profile (upper panel):  this paper ({\it open
symbols}), \citet{battaglia05} ({\it red diamonds}), and \citet{xue08} ({\it blue
hexagons}).  Number distribution with radius (lower panel, same symbols).}
 \end{figure}

\subsection{Comparison with Previous Results}

	There are few measurements of the Milky Way velocity dispersion profile at
large $R$.  The most comparable measurements are \citet{battaglia05} and
\citet{xue08}.  \citet{battaglia05} measure the velocity dispersion profile based on
a sample of 9 satellite galaxies, 44 globular clusters, 58 red giants, and 130 BHB
stars that span $10<R<140$ kpc.  \citet{xue08} measure the velocity dispersion
profile based on a sample of 2401 BHB stars that span $5<R<60$ kpc.  With the
exception of a handful of shared BHB stars, the samples are independent of one 
another.

	Figure \ref{fig:compare} compares the velocity dispersion profile and radial
number distribution of our sample with those of \citet{battaglia05} and
\citet{xue08}.  Beyond $R>50$ kpc, our sample contains a factor of 8 more stars
than \citet{battaglia05} and a factor of 2 more stars than \citet{xue08}.  
Remarkably, the velocity dispersions measured by the three samples are consistent at
the 1.5-$\sigma$ level.  The three samples also observe a statistically similar
decline in velocity dispersion with distance.  The major difference between the
samples is the significance of the observed decline in velocity dispersion with
distance.


	A linear least squares fit to the \citet{xue08} data gives a $-0.15 \pm
0.14$ \kms\ kpc$^{-1}$ decline in velocity dispersion over $10<R<60$ kpc.  The
linear fit has a standard deviation of 6.3 \kms\ and a reduced $\chi^2$ of 1.5.  A
fixed velocity dispersion has a very similar standard deviation of 6.4 \kms\ and
reduced $\chi^2$ of 1.9.  Thus the \citet{xue08} measurements cannot formally
discriminate between a constant velocity dispersion and a declining velocity
dispersion.

	The \citet{battaglia05} data, on the other hand, exhibit a steeper $-0.58
\pm 0.11$ \kms\ kpc$^{-1}$ decline in velocity dispersion over $10<R<100$ kpc.  We
exclude their last bin because it contains only 3 objects.  The linear fit has a
standard deviation of 8.6 \kms\ and a reduced $\chi^2$ of 0.85.

	The weighted average of all the samples yields a $-0.38\pm0.12$ \kms\
kpc$^{-1}$ decline in velocity dispersion over $10<R<100$ kpc.  We obtain the same
result if we average only our own two velocity dispersion measurements.  The Milky
Way velocity dispersion profile is not linear, of course, but three independent sets
of observations are in statistical agreement:  the Milky Way radial velocity
dispersion drops from $\sigma\simeq110$ \kms\ at $R=15$ kpc to $\sigma\simeq85$
\kms\ at $R=80$ kpc.

\section{CONCLUSIONS}


	The mass and mass distribution of the Milky Way are fundamental parameters
because they link directly to theoretical models.
	We use a spectroscopic sample of 910 halo stars derived from our HVS survey 
to measure the velocity dispersion profile of the Milky Way.  The stars 
are 74\% BHB stars and 26\% blue stragglers.  We estimate luminosities using
stellar evolution tracks for metal poor main sequence stars and BHB stars.  Because 
of the non-Gaussian distribution of luminosity estimates, we use a Monte Carlo 
approach to calculate velocity dispersion and its uncertainty.

	We calculate the velocity dispersion profile in two ways:  a parametric
method based on a $v_{esc}(R)$ model, and a non-parametric method based on the
caustic technique originally developed to measure galaxy cluster mass profiles.
	Comparing the two methods provides a measure of the systematic uncertainty
arising from the clipping of outliers in velocity.
	The velocity dispersion from the caustic method is 10\% smaller than the
$v_{esc}(R)$ method, but both methods identify a similar decline in velocity
dispersion with distance:  $-0.38\pm0.12$ \kms\ kpc$^{-1}$, valid for $15<R<75$ kpc.

	Our sample contains a factor of 8 more stars than \citet{battaglia05} and a
factor of 2 more stars than \citet{xue08} at $R>50$ kpc.  The velocity dispersion
profiles observed by these independent datasets are consistent at the 1.5-$\sigma$
level, and have an average velocity dispersion slope identical to our result.  
Remarkably, no matter what tracers are used, observers find the same halo velocity
dispersion profile.

	The velocity dispersion profile is a basis for measuring the total mass and
mass distribution of the Milky Way halo.
	A companion paper by Gnedin et al.\ (in preparation) presents the
theoretical calculations that turn the observed velocity dispersion profile into a
mass determination of the Milky Way.

	For further progress in measuring the Milky Way velocity dispersion profile,
it is essential to identify tracers at distances $R>50$ kpc.
	It is difficult to find $R>50$ kpc tracers because of the steep decline in
the density of the stellar halo.  It is also very difficult for proper motion 
surveys to measure tracers at $R>50$ kpc distances.
	On-going spectroscopic radial velocity surveys, such as the SDSS-3 survey
and our own HVS survey, promise to better trace the Milky Way in coming years.

\acknowledgements

	We thank M.\ Alegria, J.\ McAfee, and A.\ Milone for their assistance with
observations obtained at the MMT Observatory, a joint facility of the Smithsonian
Institution and the University of Arizona.  We also thank the referee and Oleg
Gnedin for helpful comments that improved this paper. This project makes use of data
products from the Sloan Digital Sky Survey, which is managed by the Astrophysical
Research Consortium for the Participating Institutions.  This research makes use of
NASA's Astrophysics Data System Bibliographic Services.  AD gratefully acknowledges
partial support from INFN grant PD51.  This work was supported by the Smithsonian
Institution.

{\it Facilities:} {MMT (Blue Channel Spectrograph)}

\appendix
\section{DATA TABLE}

	Table \ref{tab:dat} presents the 910 stars used here.  We provide the
observed positions, magnitudes, and velocities plus our derived luminosities,
distances, and BHB likelihood.  The table columns are:
	(1) RA (J2000), (2) Dec (J2000), (3) dereddened SDSS $g_0$ magnitude, (4)
magnitude error, (5) dereddened SDSS $(u-g)_0$ color, (6) $(u-g)_0$ error, (7)
dereddened SDSS $(g-r)_0$ color, (8) $(g-r)_0$ error, (9) heliocentric radial
velocity $v_{helio}$, (10) velocity error, (11) Galactic longitude $l$, (12)
Galactic latitude $b$, (13) Galactic rest frame velocity $v_{rf}$, as defined in
Equation \ref{eqn:vrf}, (14) BHB absolute magnitude $M_{g,BHB}$ derived from
\citet{dotter08} for [Fe/H]=$-1.9$, (15) magnitude error, (16) BHB Galactocentric
distance $R_{BHB}$, (17) blue straggler absolute magnitude $M_{g,BS}$ derived from
\citet{girardi04} for [Fe/H]=$-1.9$, (18) magnitude error, (19) blue straggler
Galactocentric distance $R_{BS}$, and (20) the star's likelihood of being BHB,
$0<f_{BHB}<1$, based on stellar colors and spectra.
	Table \ref{tab:dat} is available in its entirety in machine-readable form in
the online journal.  A portion of the table is shown here for guidance regarding its
form and content.

\begin{deluxetable}{rrcccccccccccccccccc}           
\tabletypesize{\tiny}
\tablewidth{0pt}
\tablecaption{DATA TABLE\label{tab:dat}}
\tablecolumns{20}
\tablehead{
  \colhead{RA} & \colhead{Dec} & \colhead{$g_0$} & \colhead{$\sigma_g$} &
  \colhead{$(u-g)_0$} & \colhead{$\sigma_{(u-g)}$} &
  \colhead{$(g-r)_0$} & \colhead{$\sigma_{(g-r)}$} &
  \colhead{$v_{helio}$} & \colhead{$\sigma_v$} &
  \colhead{$l$} & \colhead{$b$} & \colhead{$v_{rf}$} &
  \colhead{$M_{g,BHB}$} & \colhead{$\sigma_{BHB}$} & \colhead{$R_{BHB}$} &
  \colhead{$M_{g,BS}$} & \colhead{$\sigma_{BS}$} & \colhead{$R_{BS}$} &
  \colhead{$f_{BHB}$} \\
  \colhead{hrs} & \colhead{deg} & \colhead{mag} & \colhead{mag} &
  \colhead{mag} & \colhead{mag} &
  \colhead{mag} & \colhead{mag} &
  \colhead{\kms} & \colhead{\kms} &
  \colhead{deg} & \colhead{deg} & \colhead{\kms} &
  \colhead{mag} & \colhead{mag} & \colhead{kpc} &
  \colhead{mag} & \colhead{mag} & \colhead{kpc} &
  \colhead{} \\
  \colhead{(1)} & \colhead{(2)} & \colhead{(3)} & \colhead{(4)} &
  \colhead{(5)} & \colhead{(6)} & \colhead{(7)} & \colhead{(8)} &
  \colhead{(9)} & \colhead{(10)} & \colhead{(11)} & \colhead{(12)} &
  \colhead{(13)} & \colhead{(14)} & \colhead{(15)} & \colhead{(16)} &
  \colhead{(17)} & \colhead{(18)} & \colhead{(19)} & \colhead{(20)} \\
}
	\startdata
 0:02:05.713 &  31:18:50.23 & 20.434 & 0.026 & 0.778 & 0.095 & -0.277 & 0.040 & -209.3 & 14.0 & 110.717 & -30.385 &  -34.3 & 1.17 & 0.24 & 68.7 & 2.19 & 0.33 & 42.2 & 0.70 \\
 0:02:33.817 &  -9:57:06.85 & 18.434 & 0.021 & 0.753 & 0.040 & -0.328 & 0.040 &  -87.7 &  9.7 &  86.763 & -69.316 &  -14.8 & 1.31 & 0.12 & 33.9 & 2.06 & 0.24 & 26.2 & 0.92 \\
 0:04:36.491 &  -9:57:19.48 & 19.834 & 0.018 & 0.922 & 0.103 & -0.167 & 0.032 & -172.6 & 35.0 &  87.977 & -69.584 & -100.7 & 0.69 & 0.17 & 75.3 & 2.83 & 0.27 & 33.2 & 0.33 \\
 0:05:28.141 & -11:00:10.07 & 19.141 & 0.042 & 1.007 & 0.081 & -0.275 & 0.047 & -115.9 & 10.9 &  86.890 & -70.586 &  -47.8 & 0.94 & 0.13 & 44.6 & 2.51 & 0.19 & 22.9 & 0.64 \\
 0:07:52.013 &  -9:19:54.32 & 17.302 & 0.017 & 1.016 & 0.036 & -0.276 & 0.039 & -114.5 &  9.9 &  90.855 & -69.443 &  -42.2 & 0.91 & 0.11 & 26.3 & 2.53 & 0.16 & 16.5 & 1.00 \\
 0:12:26.890 & -10:47:54.56 & 18.898 & 0.025 & 1.006 & 0.064 & -0.321 & 0.038 & -128.2 & 10.5 &  91.741 & -71.269 &  -62.8 & 0.94 & 0.12 & 42.4 & 2.52 & 0.18 & 22.9 & 0.67 \\
 0:23:53.294 &  -1:04:46.40 & 18.199 & 0.016 & 0.749 & 0.042 & -0.255 & 0.025 &   19.7 &  9.8 & 107.552 & -63.125 &  109.0 & 1.15 & 0.17 & 30.7 & 2.22 & 0.26 & 21.1 & 0.82 \\
 0:29:31.158 &  15:39:40.20 & 19.069 & 0.024 & 1.057 & 0.069 & -0.270 & 0.038 &   22.3 & 35.0 & 115.201 & -46.881 &  153.4 & 0.88 & 0.12 & 51.2 & 2.56 & 0.18 & 27.8 & 0.53 \\
 0:36:40.570 & -11:11:25.02 & 17.421 & 0.018 & 0.778 & 0.028 & -0.304 & 0.031 &   32.5 &  9.8 & 109.940 & -73.689 &   84.1 & 1.26 & 0.12 & 21.6 & 2.09 & 0.22 & 16.6 & 0.89 \\
 0:39:06.749 &  24:09:05.62 & 19.348 & 0.018 & 0.948 & 0.081 & -0.255 & 0.030 & -130.0 & 35.0 & 119.332 & -38.634 &   15.0 & 0.94 & 0.14 & 41.4 & 2.49 & 0.21 & 17.3 & 0.55 \\
	\enddata
\tablecomments{Table \ref{tab:dat} is presented in its entirety in the
electronic edition of the Astrophysical Journal.  A portion is shown here for
guidance and content.}
 \end{deluxetable}

\clearpage


\end{document}